\documentclass[authoryear]{elsarticle}
\usepackage[english]{babel}

\usepackage[margin=0.6in]{geometry}

\usepackage[utf8]{inputenc}
\usepackage[bottom]{footmisc} 
\usepackage{graphicx}
\usepackage{rotating,booktabs}
\usepackage{natbib}
\usepackage{url}
\usepackage{booktabs}
\usepackage{graphicx}
\usepackage{adjustbox}

\usepackage{amsmath}
\usepackage{xcolor}
\usepackage{amsfonts}
\usepackage{amssymb}
\usepackage{graphicx}
\usepackage{subfigure}
\usepackage{pdflscape}
\usepackage{longtable}
\usepackage[small, justification=justified,singlelinecheck=false]{caption}
\usepackage{multirow}

\usepackage[figurename=Figure.,
justification=RaggedRight, 
labelfont={bf, footnotesize}, 
textfont={footnotesize},position=top]{caption}
\usepackage{color}
\definecolor{byzantium}{rgb}{0.44, 0.16, 0.39}	
\usepackage[english]{babel}
\usepackage{ulem}

\usepackage[margin=0.6in]{geometry}

\usepackage[utf8]{inputenc}
\usepackage[bottom]{footmisc} 
\usepackage{graphicx}
\usepackage{rotating,booktabs}
\usepackage{natbib}
\usepackage{url}
\usepackage{adjustbox}

\usepackage{amsmath}
\usepackage{amsfonts}
\usepackage{amssymb}
\usepackage{graphicx}
\usepackage{subfigure}
\usepackage{pdflscape}
\usepackage{longtable}
\usepackage[justification=justified,singlelinecheck=false]{caption}
\usepackage{multirow}
\usepackage[figurename=Figure.,
justification=RaggedRight, 
labelfont={bf, footnotesize}, 
textfont={footnotesize},position=top]{caption}
\usepackage{color}
\definecolor{byzantium}{rgb}{0.44, 0.16, 0.39}	
\usepackage{booktabs}
\usepackage{epstopdf}
\usepackage{setspace}
\usepackage[colorlinks=true]{hyperref}
\hypersetup{linkcolor=blue}
\hypersetup{citecolor=blue}
\usepackage[capposition=top]{floatrow}
\usepackage{setspace}
\usepackage{natbib}
\begin{document}

\begin{frontmatter}
\title{FTX's downfall and Binance's consolidation: the fragility of Centralized Digital Finance}

\author[myaddress,myaddress2,myaddress4]{David Vidal-Tom{\'a}s}		

\author[myaddress,myaddress2]{Antonio Briola\corref{mycorrespondingauthor}}
\ead{antonio.briola.20@ucl.ac.uk}
\cortext[mycorrespondingauthor]{Corresponding author}

\author[myaddress,myaddress2,myaddress3]{Tomaso Aste}

\address[myaddress]{Department of Computer Science, University College London, Gower Street, WC1E 6EA - London, United Kingdom.}
\address[myaddress2]{UCL Centre for Blockchain Technologies, London, United Kingdom.}
\address[myaddress3]{Systemic Risk Centre, London School of Economics, London, United Kingdom.}
\address[myaddress4]{Department of Economics, Universitat Jaume I, Campus del Riu Sec, 12071 - Castellón, Spain.}

\begin{abstract}

This paper investigates the causes of the FTX digital currency exchange's failure in November 2022. We identify the collapse of the Terra-Luna ecosystem as the pivotal event that triggered a significant decrease in the exchange's liquidity. Analysing on-chain data, we report that FTX heavily relied on leveraging and misusing its native token, FTT, and we show how this behaviour exacerbated the company's fragile financial situation. To gain further insights into the downfall, we study evolutionary dependency structures of $199$ cryptocurrencies on an hourly basis, and we investigate public trades at the time of the events. Results suggest that the collapse was actively accelerated by Binance tweets causing a systemic reaction in the cryptocurrency market. Finally, identifying the actors who mostly benefited from the FTX's collapse and highlighting a generalised trend toward centralisation in the crypto space, we emphasise the importance of genuinely decentralised finance for a transparent, future digital economy.

\hfill

	
\end{abstract}
	

\end{frontmatter}

\pagebreak

\textit{``Never in my career have I seen such a complete failure of corporate controls and such a complete absence of trustworthy financial information as occurred here. From compromised systems integrity and faulty regulatory oversight abroad, to the concentration of control in the hands of a very small group of inexperienced, unsophisticated and potentially compromised individuals, this situation is unprecedented."} John J. Ray III, Declaration in support of chapter 11 petitions and first day pleadings.
	
\section{Introduction} \label{sec:Introduction} 
In the past few years, the cryptocurrency market has experienced substantial growth, surpassing the three trillion dollar mark by the end of 2021 and still accounting for about a trillion in the present `winter' (\citealp{Statista}). The interest in cryptocurrencies has been driven by heterogeneous factors and originated a relatively mature market with competing forces. The cryptocurrency market originally provided a platform for experimenting with new financial models characterised by the potential to be decentralised and free from authority's control. However, in the last few years, we are witnessing the growth of large crypto entities operating as classic financial actors (i.e. exchanges, banks, $\dots$) while being characterised by opaque financial conditions and poor governance. This development is a cause for concern, especially considering that such unregulated centralised entities have dominant positions in the market. This new scenario contradicts the principles of transparency, independence, and accountability originally envisioned for the crypto movement.

In May 2022, the Terra-Luna stablecoin collapsed, provoking a contagion across different crypto ecosystems with long-run effects. 
As described by \cite{Briola2023}, Terra-Luna was an algorithmic stablecoin whose underlying protocol relied on a two-coin system that was not backed by traditional collaterals. Its failure was presumably induced by a liquidity pool attack and eased by the inappropriate underlying blockchain framework. This collapse remarkably damaged the confidence in the crypto market, accelerating the onset of a ``crypto winter". Users started massively withdrawing their funds from crypto institutions while investors recalled loans with cryptocurrencies as collateral. Consequently, summer 2022 was characterised by the bankruptcy of many prominent actors with excessive leverage, such as Three Arrows Capital (3AC), a Singapore-based cryptocurrency hedge fund (\citealp{Jha2022}), Voyager Digital, a cryptocurrency brokerage company (\citealp{Andersen2022}), and Celsius, a cryptocurrency lending company (\citealp{Newar2022}).

\begin{figure}[h]
	\centering	
	\caption{Schematic depiction of the leverage mechanism used by FTX and Alameda Research.}
	\begin{center}
		\includegraphics[scale=0.25]{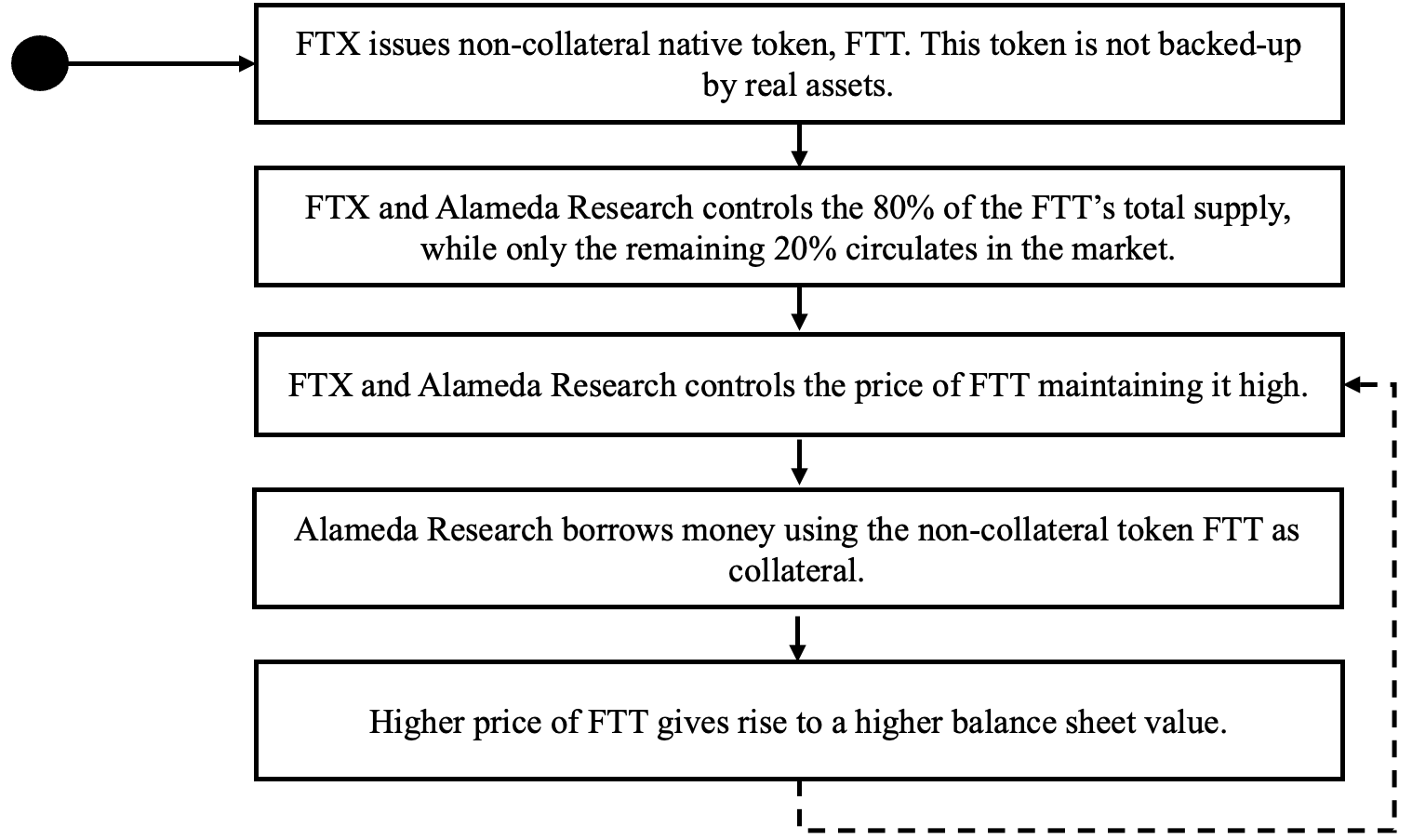}
	\end{center}
	\label{logicalflow_1}
\end{figure}

Compared to the entities mentioned above, FTX, the third-largest digital currency exchange with $\$10$ billion active trading volume and $32\$$ billion valuation at the time of events (\citealp{Fu2022, Exchanges_Cap}), was able to hide its financial situation until 02 November 2022. On this day, CoinDesk reported that Alameda Research owned $\$6$ billion FTX Tokens (FTT) in its balance sheet (\citealp{Allison2022}). In other words, the balance sheet of the leading FTX trading firm mainly included the non-collateral native token created by FTX itself. This in-house token was costless for the issuer since it was not backed by any real asset. It is worth noting that native tokens are common in centralised digital currency exchanges such as Binance (Binance token - BNB), Huobi (Huobi token - HT), and Hxro (Hxro token - HXRO). They serve as utility tokens and offer customers various incentives, including reduced trading fees, among other non-financial perks. However, the case of FTX and its token FTT concealed a deeper underlying truth. Since its Initial Coin Offering (ICO) in 2019, most of FTTs ($80\%$ of the total supply) were held by FTX and Alameda Research (\citealp{Khoo2022}). In this scenario, both entities could have easily controlled the price of the non-collateral native token, FTT, to secure additional financing while increasing the value of their balance sheets. Their financial strategy relied on a leverage mechanism where a native token without inherent value was used as collateral to raise funds. Unfortunately, this vicious cycle (see Figure \ref{logicalflow_1}) was fragile and highly exposed to external events affecting the price of FTT. As we show in this study, the Terra-Luna collapse represented such a shock. After that, both FTX and Alameda Research suffered from a credit crunch. They were initially able to avoid bankruptcy, given the misappropriation of clients' deposits, the sale of their reserves and the inflated value of FTT in their balance sheets. However, CoinDesk's report on the reliance of Alameda Research and FTX on their proprietary token unfolded the leverage mechanism used by the two companies. In response to this news, on 06 November 2022, Binance announced the liquidation of all the FTTs on its books, giving rise to a Twitter debate with FTX and Alameda Research which ended with the bankruptcy of FTX on 11 November 2022. 

The rest of the paper is organised as follows: in Section \ref{data}, we present the FTX's downfall timeline providing quantitative insights based on three heterogeneous data sources (i.e. hourly closing price, on-chain data and public transaction data). In Section \ref{methodology}, we present state-of-the-art instruments from network science used to model evolutionary dependency structures among a set of $199$ cryptocurrencies at the time of the events. In Section \ref{results_network}, we present obtained results analysing the impact of the FTX collapse on the cryptocurrency market. In Section \ref{results_herding}, we discuss the consequences of the crash, identifying the actors who mostly benefited from the FTX's collapse and highlighting a generalised trend toward centralisation in the crypto space. In Section \ref{im_fr}, we discuss the meaning of our findings, highlighting the most alarming aspects of the events presented in the paper.

\section{Data and quantitative nature of the events}\label{data}

\subsection{Hourly data analysis} \label{OHLCV_analysis}

In this paper, we use hourly USD closing prices for 199 cryptocurrencies (see the Appendix for the full list) from 01 January 2022 to 01 December 2022. The dataset is directly obtained from Binance, the largest digital currency exchange in terms of traded volume (\citealp{Exchanges_Cap}), through the use of the CCXT Python package (\citealp{Ccxt2023}).\footnote{As reported by \cite{Alexander2020} and \cite{VidalTomas2022a}, using traded data from liquid exchanges guarantees the reliability of results.}

\begin{figure}[h!]
	\centering	
	\caption{Rescaled hourly closing prices for FTT, BNB and BTC from 01 January 2022 to 01 December 2022. Events listed in Table \ref{events} are plotted with dotted lines.}
	\begin{center}
		\includegraphics[scale=0.4]{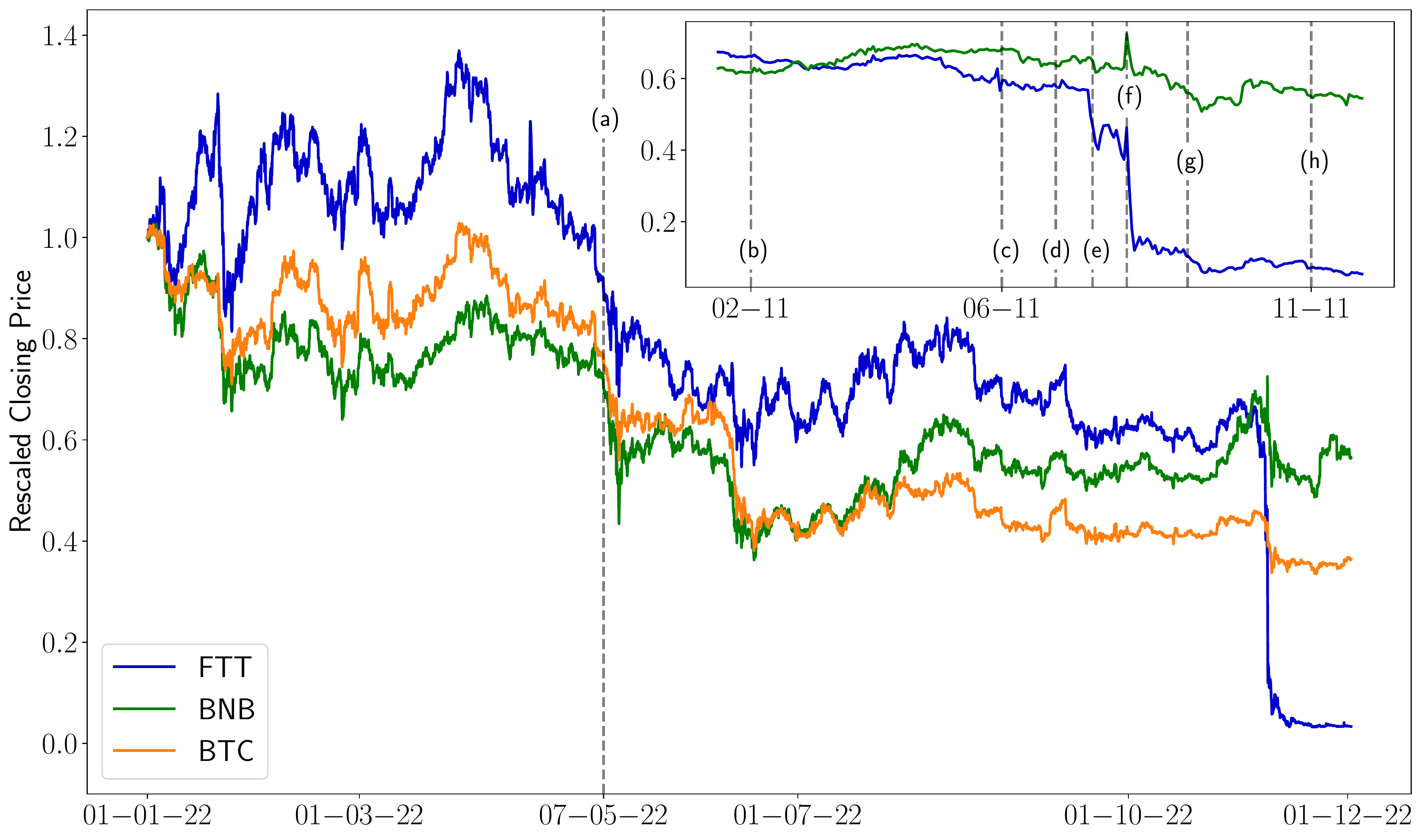}
	\end{center}
	\label{fig_des0}
\end{figure}

Figure \ref{fig_des0} reports rescaled hourly closing prices for FTT, BNB and Bitcoin (BTC).\footnote{FTT and BNB are chosen based on their role in the events analysed in the current paper, while BTC is chosen as a proxy for cryptocurrency market's behaviour.} Dotted lines highlight the main events that led to the FTX's collapse (see also \citealp{Khoo2022}, \citealp{Coghlan2022}, \citealp{Ramirez2022}, \citealp{Nathan2022}, \citealp{Conlon2022}). It is worth noting that, since 01 January 2022, FTT has demonstrated superior performance compared to BTC and BNB. As discussed in Section \ref{sec:Introduction}, such behaviour could result from a potential price manipulation of the crypto asset. However, after (a) Terra-Luna's collapse, we conjecture that FTX lost control over the FTT price due to its liquidity issues. On (b) 02 November 2022, 14:44 (GMT), CoinDesk reported that Alameda Research owned $\$6$ billion FTTs in its balance sheet (\citealp{Allison2022}). On (c) 06 November 2022, 15:47 (GMT), Binance CEO Changpeng ``CZ" Zhao announced that any remaining FTT on the company's books would have been liquidated. 

Immediately after, at 16:03 (GMT), the Alameda Research CEO, Caroline Ellison, reacted by tweeting that FTX would have bought FTTs from Binance at $\$22$ each. Due to the huge concerns throughout the crypto space regarding the financial viability of FTX and Alameda Research, Bankman-Fried (i.e. FTX's CEO) tweeted on (d) 07 November 2022, 12:38 (GMT), ``\textit{A competitor is trying to go after us with false rumours. FTX is fine. Assets are fine.}". As we show in Section \ref{transaction}, on (e) 08 November 2022, from 02:47 (GMT) to 02:48 (GMT), there was a sudden increase in selling pressure equal to $1.3$ million BUSD, which resulted in the first decrease in FTT price. In line with \cite{Khoo2022}, we conjecture that creditors recalled Alameda Research loans with FTT as collateral. Since Alameda Research could not have repaid loans, Sam Bankman-Fried was forced to ask Binance to step in and acquire the Firm. Consequently, on (f) 08 November 2022, 16:03 (GMT), Binance announced a non-binding letter of intent to purchase FTX. On (g) 09 November 2022, 15:32 (GMT), CoinDesk anticipated Binance's intention to decline any deal. Binance confirmed the leak at 20:50 (GMT) of the same day as a result of the corporate due diligence. Finally, on (h) 11 November 2022, 15:23 (GMT), FTX, and $130$ associate companies, announced that they would have commenced voluntary proceedings under chapter $11$ of the United States bankruptcy code. Table \ref{events} reports a detailed timeline of the events. 

\begin{table}[H]
\caption{Timeline of the events that led to the FTX bankruptcy in November 2022.}
\label{events}
\resizebox{\columnwidth}{!}{%
\begin{tabular}{ccc}
\hline
\textbf{Date} &
  \textbf{Reference} &
  \textbf{Event} \\ \hline
07-05-2022 22:00 &
  (a) &
  Terra-Luna collapses. First day that Terra (USDT) lost the peg to USD. \\ \hline
02-11-2022 14:44 &
  (b) &
  \begin{tabular}[c]{@{}c@{}}Coindesk reported that the value of Alameda Research heavily relied on the FTX's in-house tokens, FTT. \\ Specifically, Alameda Research owned \$$14.6$ billion of assets and \$$6$ billion were FTT.\end{tabular} \\ \hline
06-11-2022 15:47 &
  (c) &
  \begin{tabular}[c]{@{}c@{}}Changpeng ``CZ" Zhao (i.e. Binance CEO) announced that his company would have liquidated any remaining FTT on Binance books. \\ In response to this announcement, at 16:03 (GMT), Caroline Ellison (i.e. Alameda Research CEO), \\ tweeted that FTX would have bought all the FTT tokens from Binance at a value of \$$22$ each.\end{tabular} \\ \hline
07-11-2022 12:38 &
  (d) &
  \begin{tabular}[c]{@{}c@{}}Due to the huge concerns throughout the crypto space regarding the financial viability of FTX and Alameda Research, \\ Bankman-Fried tweeted ``\textit{A competitor is trying to go after us with false rumors. FTX is fine. Assets are fine}".\end{tabular} \\ \hline
08-11-2022 02:48 &
  (e) &
  \begin{tabular}[c]{@{}c@{}}We identify a massive selling pressure on FTT, which could be related to the liquidation of Alameda Research loans.\end{tabular} \\ \hline
08-11-2022 16:03 &
  (f) &
  \begin{tabular}[c]{@{}c@{}} Binance announced the existence of a non-binding letter of intent to purchase FTX.\end{tabular} \\ \hline
09-11-2022 15:32 &
  (g) &
  \begin{tabular}[c]{@{}c@{}}Coindesk anticipated Binance intention to decline any kind of deal. The news was officially confirmed at 20:50 (GMT). \end{tabular} \\ \hline
11-11-2022 15:23 &
  (h) &
  \begin{tabular}[c]{@{}c@{}}FTX and its 130 related companies, announced that they commenced voluntary proceedings \\ under chapter 11 of the United Stated bankruptcy code.\end{tabular} \\ \hline
\end{tabular}%
}
\end{table}

\subsection{On-chain data analysis}\label{sec_on-chain}

To quantitatively prove the relevance of Terra-Luna's collapse on the FTX's bankruptcy, we use high-quality on-chain data from \cite{Glassnode2022}. 

After the failure of the algorithmic stablecoin, both FTX and Alameda Research suffered from a credit crunch caused by the decrease in FTT's price and the increased difficulty in obtaining credit from lenders. Indeed, bankruptcies characterising summer 2022 (i.e. 3AC, Voyager Digital and Celsius) fomented market uncertainty, decreasing lending volume and causing a generalised down-market. As a consequence, since May 2022, FTX and Alameda Research could no longer control FTT price, and the leverage mechanism described in Figure \ref{logicalflow_1} was abruptly interrupted. The Wall Street Journal reports that the CEO of Alameda Research informed her employees that the Firm used FTX clients' funds to pay back creditor's loans that were being recalled due to the credit crunch triggered by Terra-Luna collapse (\citealp{De2022}). The apology letter sent on November 2022 by Sam Bankman-Fried to his employees further confirms the crucial role of Terra-Luna's failure in FTX bankruptcy: ``\textit{I believe that the events that led to the breakdown this month} [November 2022] \textit{included a crash in markets this spring} [Terra-Luna] \textit{that led to a roughly $50\%$ reduction in the value of collateral}" (\citealp{Rooney2022}).  

\begin{figure}[H]
	\centering	
	\begin{center}
        \includegraphics[scale=0.4]{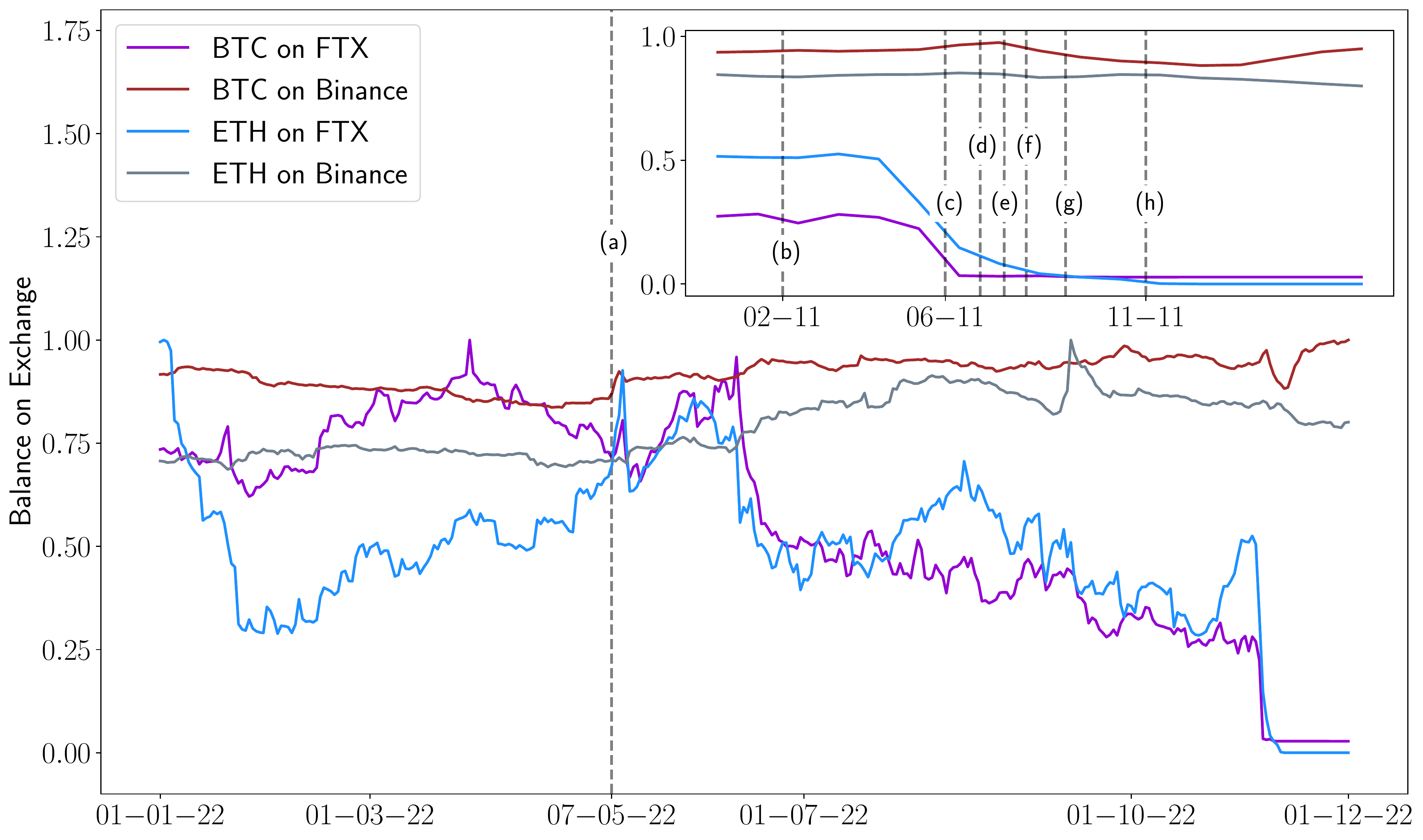}\\
	\end{center}
	\caption{Rescaled total amount of Bitcoins and Ethereum Coins held on FTX and Binance addresses from 01 January 2022 to 01 December 2022. The data are retrieved from \cite{Glassnode2022}.}
	\label{fig:on_chain_data}
\end{figure}

\begin{figure}[h!]
	\centering	
	\caption{Schematic depiction of the mechanism that led to the FTX collapse.}
	\begin{center}
		\includegraphics[scale=0.32]{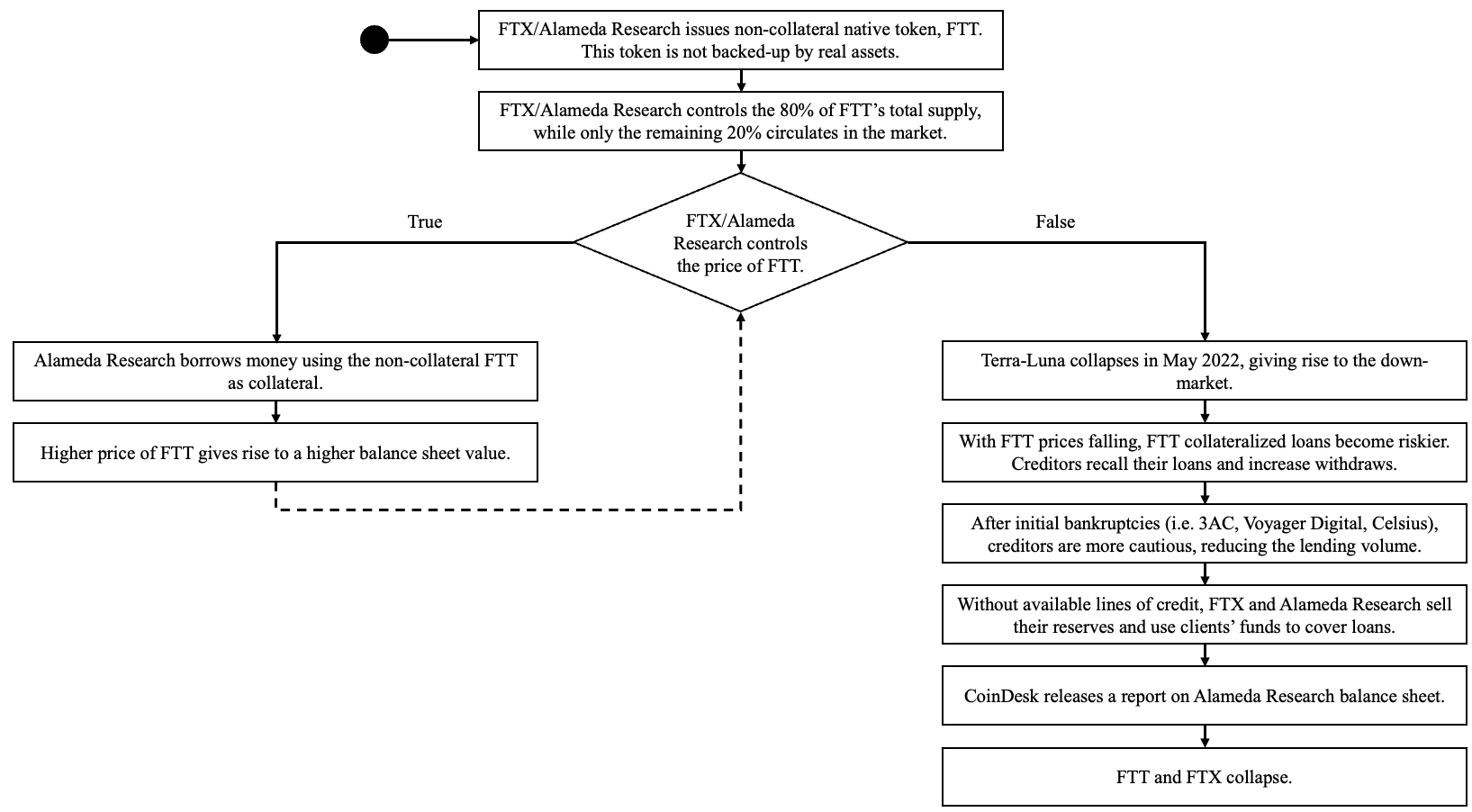}
	\end{center}
	\label{logicalflow}
\end{figure}

In order to validate our events' reconstruction, we report in Figure \ref{fig:on_chain_data} the rescaled total amount of reserves (in BTC and Ethereum) owned by FTX and Binance from 01 January 2022 to 01 December 2022. FTX, differently from its main competitor (i.e. Binance), started to sell its reserves slightly after the Terra-Luna collapse in May 2022 to overcome the credit crunch. Based on these findings, in Figure \ref{logicalflow}, we extend Figure \ref{logicalflow_1} by incorporating a new branch that depicts the circumstances that disrupted the vicious cycle involving FTX and the consequences observed since May 2022.

\subsection{Transaction data analysis}\label{transaction}
To gain further insights on the FTX's bankruptcy, we analyse FTT's public trades occurred on the Binance digital currency exchange at the time of the events. Also in this case, the dataset is directly obtained from Binance digital currency exchange using the CCXT Python package (\citealp{Ccxt2023}). Unlike hourly closing prices, transaction data are expressed in Binance USD (BUSD), as this is the primary exchange pair on Binance. 

\begin{figure}[h!]
	\centering	
	\begin{center}
		\includegraphics[scale=0.4]{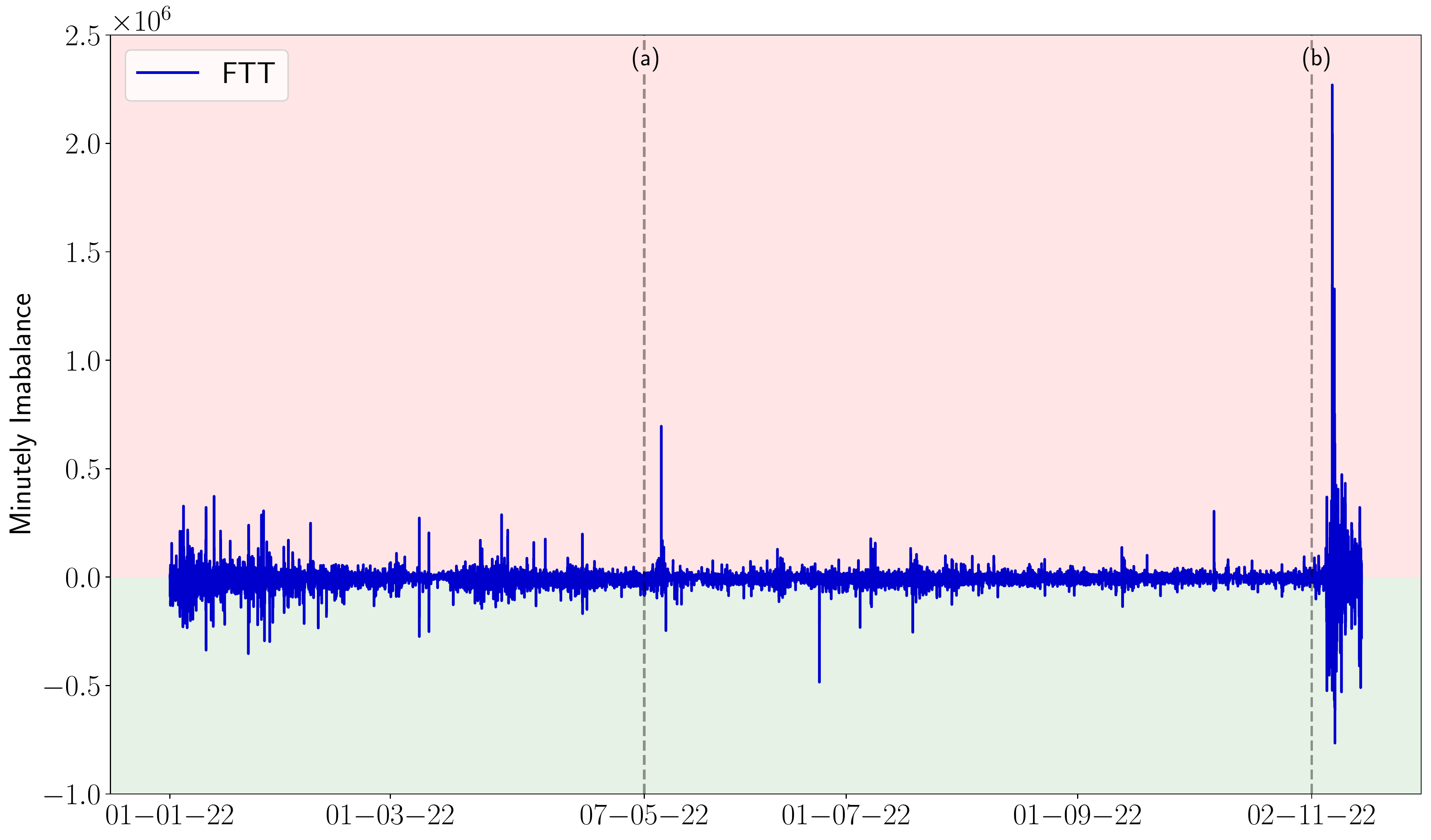} \\
		\includegraphics[scale=0.4]{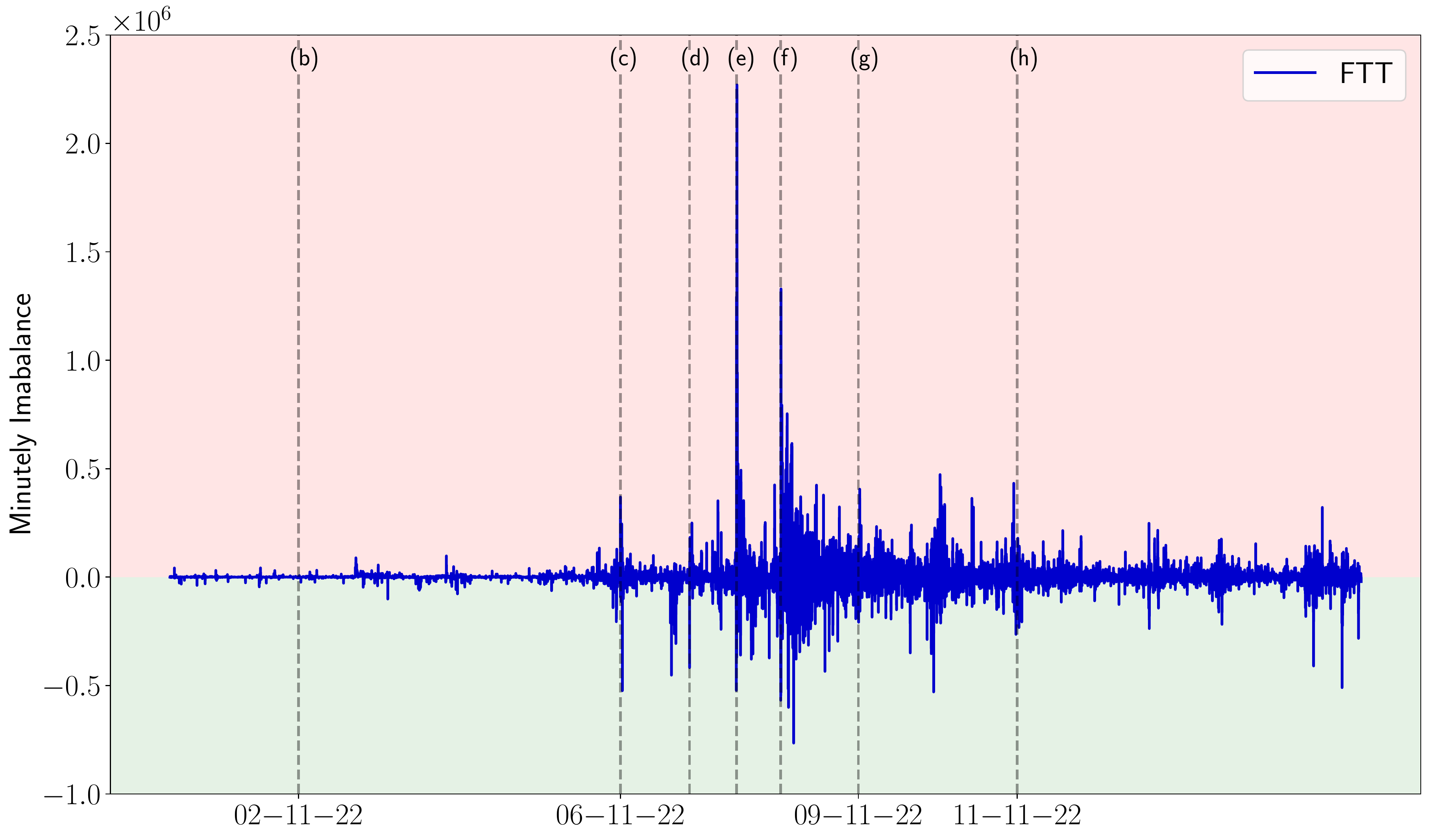}
	\end{center}
	\caption{Minutely imbalance for FTT on the Binance digital currency exchange. Positive values (red color area) and negative values (green color area) denote selling and buying pressure, respectively.}
	\label{imbalance}
\end{figure}

Figure \ref{imbalance} reports minutely imbalances, where positive values (red area in the plot) indicate a selling pressure, while negative ones (green area in the plot) indicate a buying pressure.\footnote{The calculation of the imbalance is based on the methodology described in \cite{Briola2023}. The procedure involves separating data into buy and sell trades, computing the transaction costs by multiplying the volume of each transaction by its execution price, aggregating the transaction costs by minute, and subtracting the total of buy transactions costs from the total of sell transactions costs to obtain imbalances.} Our analysis reveals that, prior to November 2022, the highest selling pressure in FTT occurred on 12 May 2022 with a value of BUSD $695,690$ as a direct consequence of the Terra-Luna's failure. This finding enforces our statement about the crucial role of the Terra-Luna's collapse in FTX's bankruptcy, as no other significant event appears to have influenced the market, including the Russia-Ukraine conflict (\citealp{BedowskaSojka2022}) and the implementation of contractionary monetary policies (\citealp{Castrovilli2022}). In Figure  \ref{imbalance}, we further observe that, in November 2022, market dynamics were affected by the Twitter debate involving FTX, Alameda Research and Binance. CoinDesk's report alone (b) had a limited impact on the market. In contrast, the first significant shock is observed on (c) 06 November 2022 at 15:47 (GMT), when the Binance CEO announced the intention to liquidate FTT reserves, leading to a rise in selling pressure at 15:49 (GMT), amounting to BUSD $369,420$. In light of the high buying pressure at 16:23 (GMT) (i.e. BUSD $523,730$), we postulate that Binance elicited a counter-reaction from FTX. Despite the efforts by the FTX's CEO to restore investors' trust (d), FTT experienced a significant selling pressure equal to BUSD $1.3$ million on (e) 08 November 2022, at 02:48 (GMT), when it was traded at BUSD $21.83$. The most significant selling pressure (i.e. BUSD $2.56$ million) is detected at 02:56 (GMT) of the same day when FTT had already dropped to BUSD $19.6$.\footnote{After Caroline Ellis announced that Alameda Research would have purchased all of Binance's FTT holdings for USD $22$ per token, FTT traded between BUSD $21.83$ (i.e. USD $21.85$) and BUSD $22$ (i.e. USD $22.01$) for a couple of hours, suggesting that USD $22$ was not a psychological barrier for investors.} In agreement with \cite{Khoo2022}, a plausible explanation could be that Alameda Research had loans to be liquidated when the price of FTT would have fallen below BUSD $21.8$. Hence, we cannot exclude that this abnormal selling pressure could have been generated by FTX itself trying to repay loans collateralized by FTT. In this scenario, having already used the majority of clients' funds and most of the reserves to front the credit crunch triggered by the Terra-Luna's collapse (see Section \ref{sec_on-chain}), FTX did not have alternative sources of liquidity. Despite its origin, this event led to the technical collapse of FTX, as observed on (f) 08 November 2022. Sam Bankman-Fried asked Binance to acquire the FTX group, further spreading panic among investors and leading to a significant drop in the FTT price (see Figure \ref{fig_des0}). On (g) 09 November 2022, Binance declined to acquire FTX, generating additional selling pressure. On (h) 11 November 2022, the bankruptcy of FTX was announced with minimal impact, as investors had already taken into account the collapse, and the price was BUSD $2.79$.

\section{Methodology}\label{methodology}

\subsection{Network analysis: Triangulated Maximally Filtered Graph (TMFG)}\label{methodology_network}
To describe the impact of FTX's collapse on the market's dynamics, we analyse the evolution of a set of $199$ cryptocurrencies' dependency structures during the crash. As shown in \cite{Briola2023}, we use the Pearson correlation coefficient to model linear relationships among the assets. It is worth noting that during periods of stress in the underlying system, pure correlations may exhibit heightened sensitivity. An exponential time-weighting structure that assigns a larger weight to the latest observations and lower weights to older observations can mitigate this effect. Results show that weighted correlations are smoother and more resilient to market turbulence than unweighted ones. Additionally, weighted correlations are more effective in distinguishing genuine correlations from spurious ones. Following the definition in \cite{Pozzi2012}, we define the Pearson correlation coefficient weighted with exponential smoothing as follows:

\begin{equation}
\label{eq:exponentially_smoothed_corr_coef}
    \rho_{i,j}^w = \frac{\sum_{t=1}^{\Delta t} w_t (y_t^i - \overline{y}_i^w)(y_t^j - \overline{y}_j^w)}{\sqrt{\sum_{t=1}^{\Delta t} w_t (y_t^i - \overline{y}_i^w)^2} \sqrt{\sum_{t=1}^{\Delta t} w_t (y_t^j - \overline{y}_j^w)^2}} .
\end{equation}

where $w_t = w_0 e^{\frac{t-\Delta t}{\theta}}, \forall t \in \{1, 2, \dots, \Delta t\} \land \theta > 0$ represents an exponentially smoothed weight structure such that $\sum_{t=1}^{\Delta t} w_t = 1$ and  $\overline{y}_k^w = \sum_{t=1}^{\Delta t} w_t y_t^k$. $\Delta t$ corresponds to rolling windows made of 24 hours with steps of 1 hour each, and $\theta$ is set to $0.1$.\footnote{The results are consistent for different values of $\theta$.} Based on Equation \ref{eq:exponentially_smoothed_corr_coef}, we use the Triangulated Maximally Filtered Graph (TMFG) (\citealp{Massara2017, briola2023topological}) to model dependencies among cryptocurrencies. This information filtering technique presents several advantages compared to alternative methods. It can capture meaningful interactions among multiple assets and exhibits topological constraints that facilitate regularisation in probabilistic modelling (\citealp{aste2022topological}). Given the system's network structure, the time-dependent influence of each asset is finally measured through the Eigenvector Centrality (\citealp{bonacich2007some}).

\subsection{Buy and hold returns}\label{methodology_bhr}

We use buy-and-hold returns (BHR) to analyse the financial performance of the cryptocurrencies in our dataset (see \citealp{Momtaz2021, briola2021deep, VidalTomas2022, vidal2023illusion}). In the current work, BHR is defined as the relative difference between prices on 01 December 2022 and 01 January 2022. This simple computation allows to quantify the investors' trust in Binance ecosystem compared to alternative ones.

\section{Results}\label{results}

\subsection{FTX's collapse: Correlations and network analysis}\label{results_network}

\begin{figure}[h!]
	\centering	
	\begin{center}
		\includegraphics[scale=0.45]{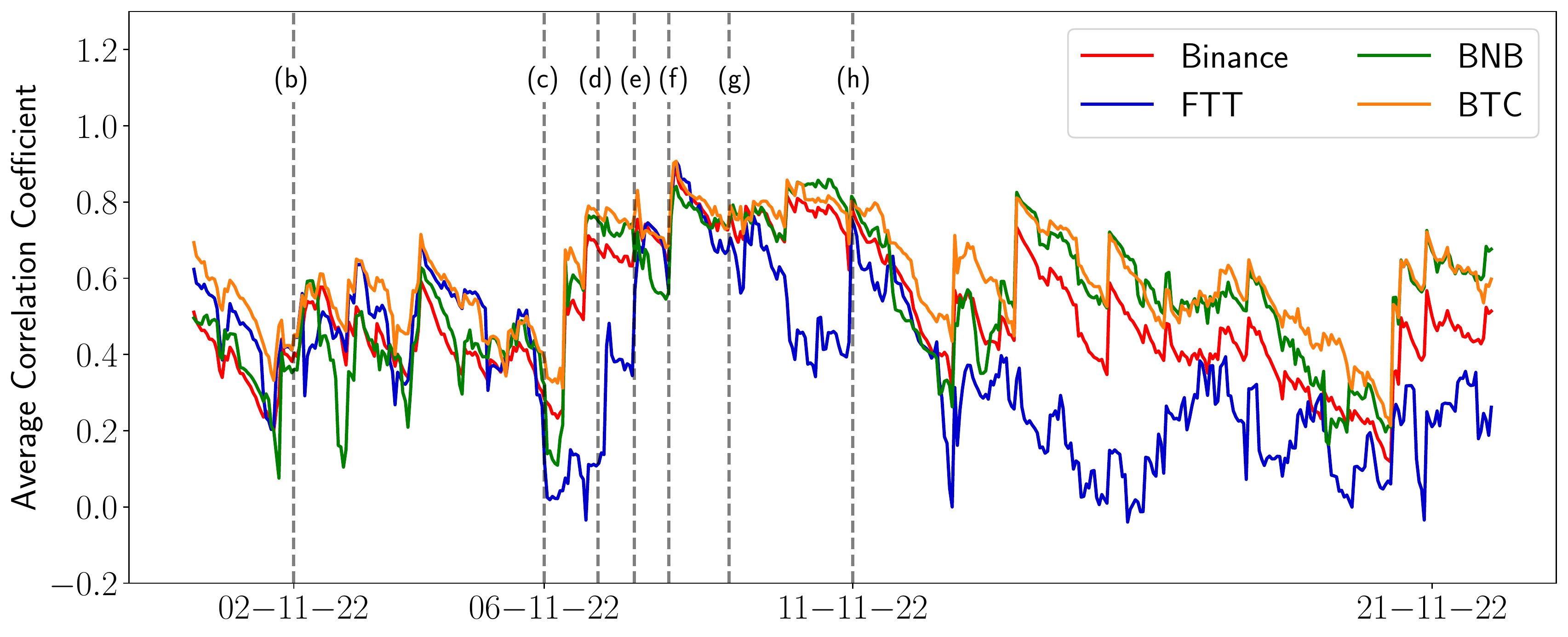}
	\end{center}
	\caption{Exponentially smoothed weighted correlations for FTT, BNB, BTC and Binance digital currency exchange, using 24 hour rolling windows with steps of 1 hour each. Binance (red line) denotes the average correlation of all the 199 cryptocurrencies, while FTT (blue line), BNB (green line) and BTC (orange line) represent their average correlation with the rest of the system.}
	\label{netfig1}
\end{figure}

Figure \ref{netfig1} reports exponentially smoothed average correlation coefficients for FTT, BNB, BTC, and the Binance digital currency exchange.\footnote{The Binance digital currency exchange represents the average correlation of all the $199$ cryptocurrencies available in the study.} In line with findings in Section \ref{transaction}, results indicate that the CoinDesk report did not significantly impact the market's dynamics. The first notable event is observed in (c) when the Binance CEO announced the intention to liquidate all the FTT reserves held by his company. The announcement gave rise to the complete disconnection of FTT from the market (Binance), with an average correlation coefficient close to $0$. This finding is coherent with results provided by \cite{Conlon2022}, who observed the first significant negative FTT's response on 06 November 2022. Afterwards, we identify a continuous increase in market correlations since the whole market reacted to the flow of FTT-related news (i.e. (d), (e) and (f)). The maximum correlation is observed on (f) 08 November 2022, at 19:00 (GMT), shortly after Binance announced a non-binding letter of intent to acquire FTX. 

\begin{figure}[h!]
	\centering	
	\begin{center}
		\includegraphics[scale=0.4]{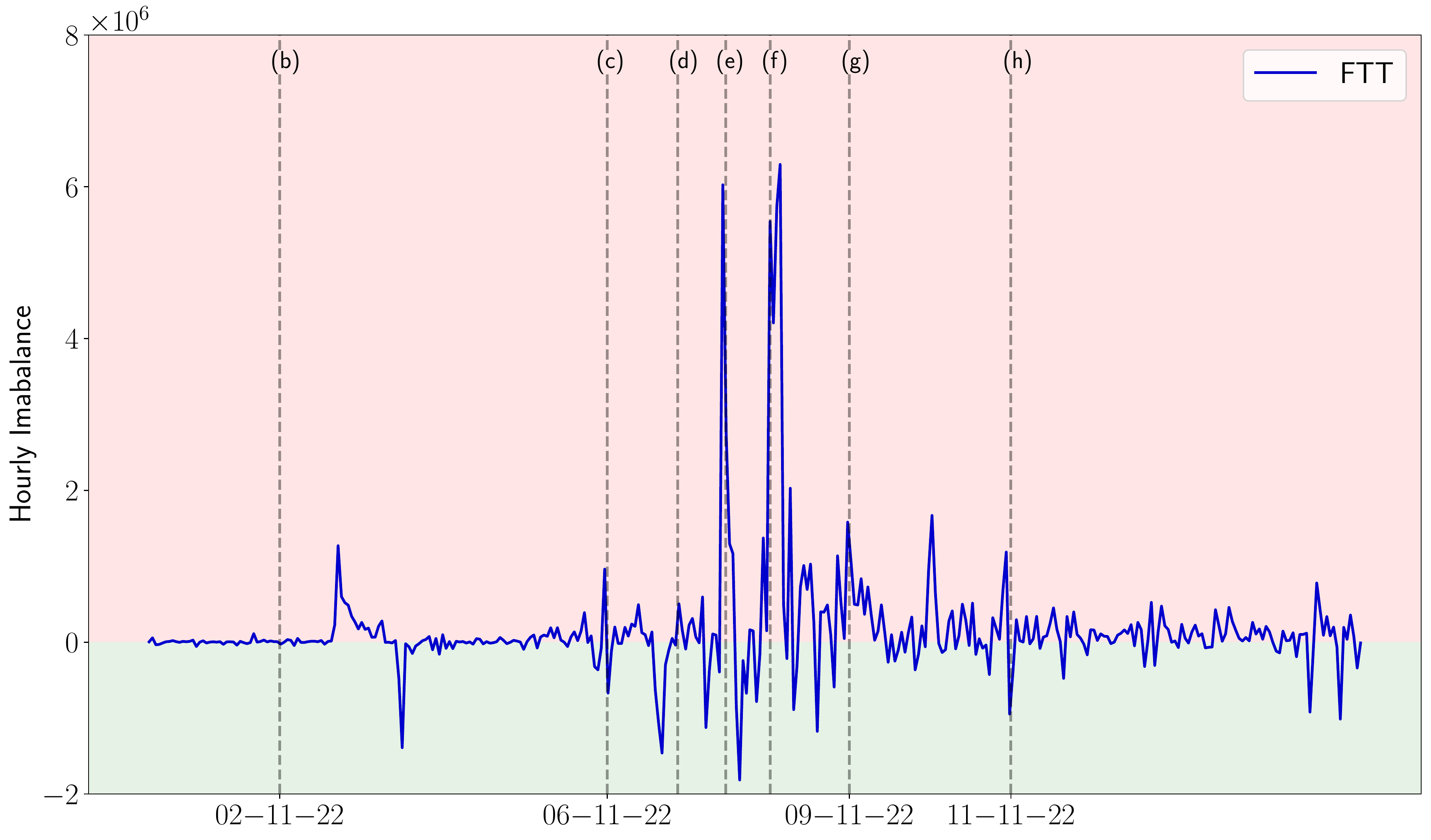}
	\end{center}
	\caption{Hourly imbalance for FTT on the Binance digital currency exchange. Positive values (red color area) and negative values (green color area) denote selling and buying pressure, respectively.}
	\label{fig:hourly_imbalance}
\end{figure}

The maximum market correlation coincides with the highest hourly selling pressure in FTT, with BUSD $6.29$ million in (net) sales (see Figure \ref{fig:hourly_imbalance}), which shows the systemic effect of the FTX's collapse on the market. The trend persisted until the official FTX's bankruptcy (h) when the market correlation decreased remarkably. FTT was then ``excluded" from the crypto system.

\begin{figure}[h!]
	\centering	
	\caption{Non-overlapping eigenvector centrality of FTT, BNB and BTC using a 24h rolling window. Color areas show the distribution of the eigenvector centrality for the rest of cryptocurrencies considering $1\%$-$99\%$, $5\%$-$95\%$, and $25\%$-$75\%$ percentiles.}
	\label{netfig2}
	\begin{center}
		\includegraphics[scale=0.45]{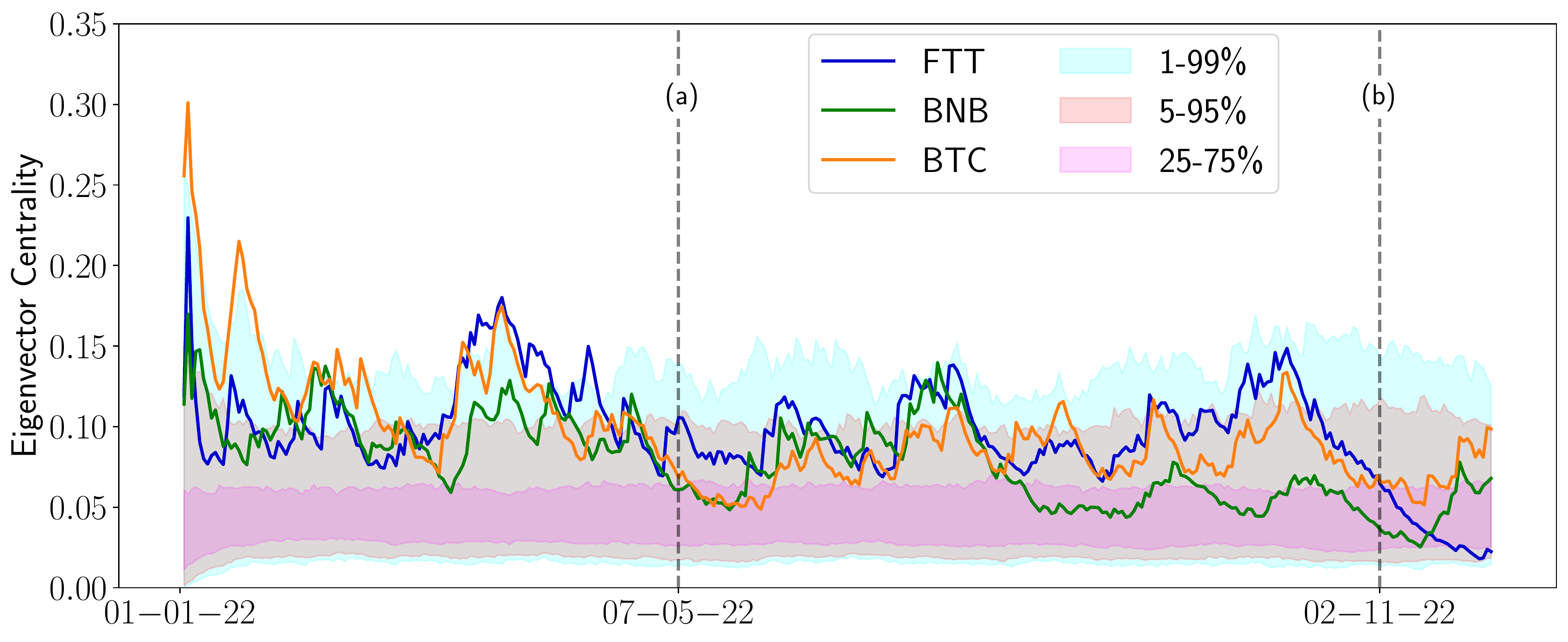}
	\end{center}
\end{figure}

To enhance our understanding of the system's collective dynamics during the FTX's downfall, we examine assets' centralities within the TMFG. Figure \ref{netfig2} illustrates the temporal evolution of the Eigenvector Centrality for FTT, BNB, and BTC, computed on non-overlapping rolling windows of $24$ hours. We highlight two interesting findings. First, the impact of CoinDesk's report (b) on the FTT Eigenvector Centrality is worth noting. Despite a lack of significant effects on prices (see Figure \ref{fig_des0}) and market correlations (see Figure \ref{netfig1}), we observe a decline in Eigenvector Centrality (see Figure \ref{netfig2}) in correspondence of this event. This suggests that following the report's publication on 02 November 2022, FTX and Alameda Research may have ceased their speculative operations with FTT.

Second, Figure \ref{netfig2} sheds additional light on the potential misuse of FTT. As a utility token, it should have been utilised by FTX to offer incentives to users, such as reduced trading fees or the ability to pay for goods and services.\footnote{See \citealp{Binance2023} for an overview of the usage of BNB, a utility token similar to FTT} This token was, therefore, not intended to be mainly used for speculative purposes. 
In other words, FTT and native tokens should be characterised by a low market correlation and degree of centrality by nature. This phenomenon was firstly analysed by \cite{Briola2022}, who found that centralised exchange tokens (e.g., BNB, HT, and HXRO) are characterised by lower market correlation (and lower centrality) compared to digital currencies (e.g., BTC and Litecoin) or smart contract tokens (e.g., ETH and Tron). In that paper, the authors also found a suspicious result, apparently without a clear explanation: FTT was characterised by a high degree of centrality, similar to the one of more speculative cryptocurrencies such as BTC and Litecoin. Given that the authors utilised data from the FTX digital currency exchange, they hypothesised that this result could have been explained by ``\textit{an overestimation of the role played by the exchange-specific token, FTT}". As depicted in Figure \ref{netfig1}, FTT exhibits a high degree of centrality also in the Binance digital currency exchange, with peaks higher than ones of the most speculative tokens. In light of what described in the current research work, we can assert, with sufficient confidence, that this behaviour was due to the misuse of FTT as a speculative currency. Specifically, users could only use the $20\%$ of the total supply as a utility token. In contrast, $80\%$ of the supply was used for speculative purposes by Alameda Research's and FTX's managers to take advantage of the upward market and drive up FTT's price. In other words, given the unbalanced FTT supply distribution, FTX's managers could have inflated the token's price during up-market periods as long as credit lines were available. This misuse was reflected in a higher correlation and centrality of FTT. On the contrary, during its ICO, BNB was better distributed among heterogeneous actors, including the foundation team ($40\%$), angel investors ($10\%$), and the general public ($50\%$) (\citealp{Cointelegraph2022}). This distribution guaranteed a fair valuation of BNB and correct use as a utility token by Binance's users, giving rise to a lower degree of centrality, as observed by \cite{Briola2022}.

\subsection{Binance: the raise of Centralised Digital Finance}\label{results_herding}

Despite the tremendous effects on the trust on the crypto movement, the failure of FTX in November 2022 was beneficial for the exchange's immediate competitor: Binance. In Table \ref{tab:best_worst_crypto}, we report the ten cryptocurrencies with the worst and best performance in terms of BHR from 01 January 2022 to 01 December 2022. The median BHR for the sample is $-79\%$, with $25^{th}$ and $75^{th}$ percentiles of $-69\%$ and $-87\%$, respectively. Despite the Ukraine-Russia conflict (\citealp{BedowskaSojka2022}), Terra-Luna's collapse (\citealp{Briola2023}) and contractive monetary policies (\citealp{Castrovilli2022}), BNB is among the best performing assets, with a BHR of $-43.5\%$. This result demonstrates the investors' confidence in Binance and the consolidation of the cryptocurrency market around this Firm. Moreover, as a consequence of the FTX's failure, Binance reported a $30\%$ increase in trading activity (\citealp{Pan2022}), further emphasising its growing dominance in the crypto space. This point is also supported by the number of daily active users in the blockchain infrastructure, on 01 December 2022, since BNB chain was the leader with $1,497,102$ daily active users, followed by Ethereum ($313,110$), Polygon ($361,252$), PancakeSwap ($146,097$) and Solana ($107,943$) (\citealp{tokenterminal2023}).

\begin{table}[H]
\small
\caption{Buy and hold (BHR) returns from 01 January 2022 to 01 December 2022. First row reports cryptocurrencies showing the most negative returns, while the second row reports cryptocurrencies showing the most positive returns.}
\label{tab:best_worst_crypto}
\resizebox{\columnwidth}{!}{%
\begin{tabular}{ccccccccccc}
\textbf{Crypto (-)} & SPELL  & RAY    & FTT    & ILV    & MOVR   & JASMY  & MIR    & PERP   & GALA   & HNT    \\ \hline
\textbf{BHR}        & -0.974 & -0.971 & -0.967 & -0.959 & -0.956 & -0.947 & -0.947 & -0.945 & -0.945 & -0.941 \\ \hline
                    &        &        &        &        &        &        &        &        &        &        \\
\textbf{Crypto (+)}          & CHZ    & BNB    & ETC    & UNFI   & DOGE   & XMR    & QNT    & TRX    & LAZIO  & TWT    \\ \hline
\textbf{BHR}        & -0.442 & -0.435 & -0.428 & -0.426 & -0.408 & -0.382 & -0.323 & -0.287 & 0.068  & 2.123  \\ \hline
\end{tabular}%
}
\end{table}

The growing dominance of Binance can be further assessed by considering the top two crypto assets performers in 2022, LAZIO and TWT. On the one hand, LAZIO appears to have a financial advantage in its niche due to the presence on Binance digital currency exchange. This advantage is further bolstered by Binance's sponsorship of S.S. Lazio football club, which prominently displays the Binance brand on the team's jerseys (\citealp{Proch2021}). On the other hand, TWT is the native token of Trust Wallet, a self-custodian cryptocurrency wallet founded by Viktor Radchenko in November 2017 and acquired by Binance in July 2018. Interestingly, on 13 November 2022, Binance CEO tweeted about the advantages of self-custodianship and the role of Trust Wallet in this regard, leading to a $47\%$ increase in the value of TWT (\citealp{Somraaj2022}).

\begin{figure}[h]
	\centering	
	\begin{center}
		\includegraphics[scale=0.45]{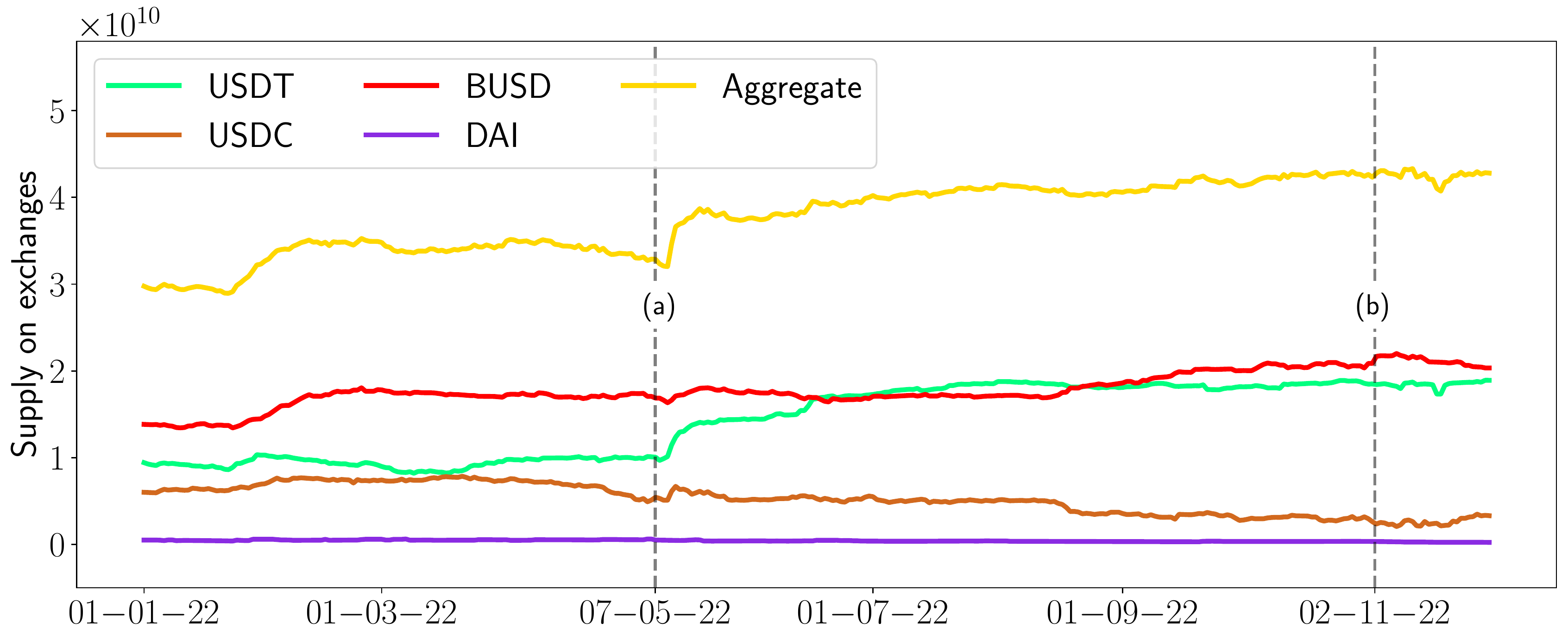}	
	\end{center}
	\caption{Supply held on exchanges' reserves of the top four stablecoins (i.e. USDT, USDC, BUSD and DAI) from 01 January 2022 to 01 December 2022. Also the corresponding aggregate value is reported. The data are retrieved from \cite{Glassnode2022}.}
	\label{stablecoins}
\end{figure}

\begin{figure}[h]
	\centering	
	\begin{center}
		\includegraphics[scale=0.45]{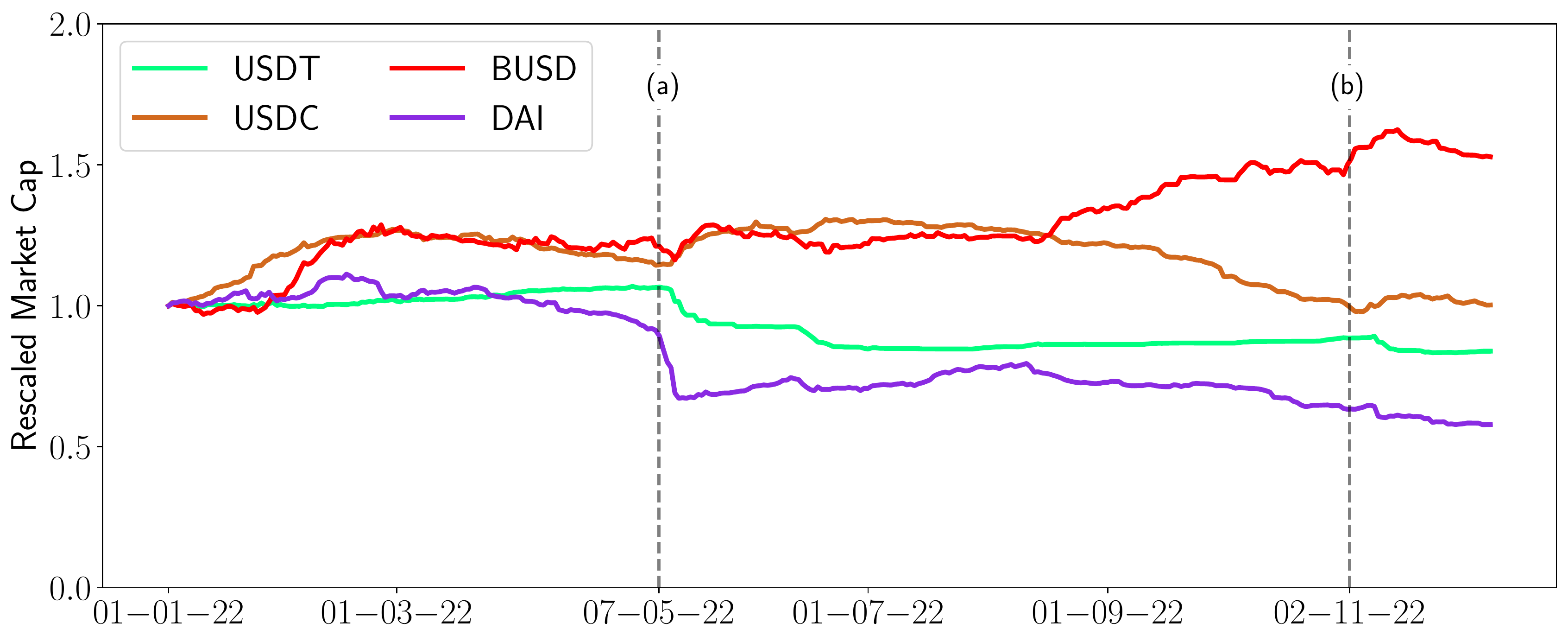}	
	\end{center}
	\caption{Rescaled market capitalisation of the top four stablecoins (i.e. USDT, USDC, BUSD and DAI) from 01 January 2022 to 01 December 2022. Data are retrieved from \cite{Defillama2023}.}
	\label{rescaledmarketcap}
\end{figure}

In line with this findings, we also highlight Binance's relevance in the stablecoin market with the presence of BUSD. As shown in Figure \ref{stablecoins}, on 01 December 2022, BUSD represented approximately the $50\%$ of the entire stablecoins' supply on digital currency exchanges. Similarly, Figure \ref{rescaledmarketcap} shows how BUSD increased its value by $54\%$ since 01 January 2022, while its main centralised competitors (i.e. USDT and USDC) registered comparatively worse market performances (i.e. $-17\%$ and $1\%$, respectively). Interestingly, the decentralised option, DAI, was the most affected by the Terra-Luna's collapse, with a decrease in market cap equal to  $-42\%$. This result could highlight a potential shift in users' sentiment, with the centralized option preferred over the decentralized one. This would be in line with what is stated in \cite{Duan2023}, where the authors observe that ``\textit{BUSD is found as the most stable stablecoin with the fastest correction speed}", while DAI is the least stable.\footnote{BUSD is a Binance branded stablecoin, issued by Paxos Trust comapny. Paxos is a regulated institution supervised by the New York Department of Financial Services (NYDFS). On 13 February 2023, NYDFS ordered Paxos Trust to stop the issuance of BUSD, since the United States Securities and Exchange Commission alleged that BUSD is an unregistered security (\citealp{Partz2023}). Binance informed that they ``\textit{will make product adjustments accordingly}" (\citealp{Zhao2023}).}

\section{Conclusion}\label{im_fr}
This paper investigates the causes and effects of the FTX's failure in November 2022. Our contribution to the existing literature is threefold. First, we use three different data sources (i.e. hourly closing prices, on-chain data and transaction data) to quantitatively analyse the events at different granularities. Second, we study the evolution of dependency structures among $199$ cryptocurrencies and capture the phases of the market's reaction to the ongoing downfall. We prove how the absence of regulation and the lack of transparency allowed FTX to build a leverage mechanism characterised by (i) the issuance of non-collateralised native tokens (FTT), (ii) control over the majority of FTTs, and (iii) unlimited loan requests using FTT as collateral despite its lack of inherent value. We also show that the decline of FTX was triggered by Terra-Luna's crash, which resulted in a decrease in FTT's price and a sudden reduction of credit availability. Despite the attempts to hide the compromised financial situation by selling digital reserves and misappropriating customers' funds to pay loans, the reliance of FTX and Alameda Research on FTT was finally reported by CoinDesk, raising a Twitter debate on the stability of the Firm. We identify the Binance announcement to sell FTT reserves as the catalyst for FTX's collapse. At the same time, the systemic impact of the downfall on the cryptocurrency market was evident only after the attempt to sell the company. As a third contribution, we analyse the effects of the FTX's collapse on the process which is driving the crypto movement toward centralisation. Specifically, we demonstrate that the consolidation of Binance's leading role in the crypto space in response to the FTX's downfall recalls the urgency to protect users preventing the creation of opaque monopolies. In 2022, Binance's volume market share increased from $48.7\%$, in the first quarter, to $66.7\%$, in the last quarter  (\citealp{Cryptocompare_review}). When Bitcoin was created in 2008, Satoshi Nakamoto (\citealp{Nakamoto2008}) stated that "\textit{what is needed is an electronic payment system based on cryptographic proof instead of trust, allowing any two willing parties to transact directly with each other without the need for a trusted third party}". After $15$ years, the cryptocurrency industry appears to be moving towards centralisation, with third-party entities serving as the primary means for exchanging cryptocurrencies. Despite not being created by central banks, cryptocurrencies are now predominantly managed by unregulated private companies acting as traditional financial institutions (e.g. paying interests for deposits, providing landings and releasing debit cards). These centralised and unregulated entities cannot be considered part of the new digital economy since they are a transposition of the existing regulated financial institutions inside the crypto space. Decentralised Finance (DeFi) should be the obvious candidate to support the future digital economy, given that it naturally provides users on-chain transparency, self-custody, governance, and fair access to financial products. Consequently, DeFi could avoid the governance issues represented by FTX, whose managers were able to raise USD $2$ billion from $80$ investors (\citealp{Griffith2022}), misuse users' funds, and create an articulate corpus of $130$ side companies without any supervision. Unfortunately, as underlined by \cite{Fu2022} and \cite{Aramonte2021}, given its security risks and excessive concentration of decision power in the hands of large coin-holders, DeFi cannot yet be considered a mature solution.

\section*{Acknowledgements}

The author, D.V-T., acknowledges the financial support from the Margarita Salas contract MGS/2021/13 (UP2021-021) financed by the European Union-NextGenerationEU. The author, T.A, acknowledges the financial support from ESRC (ES/K002309/1), EPSRC (EP/P031730/1) and EC (H2020-ICT-2018-2 825215).

\bibliography{bibliography}

\pagebreak

\section*{Supplementary Material}\label{appendix: Supplementary_Material}

\begin{longtable}[c]{@{}cccc@{}}
\caption*{List of the $199$ cryptocurrencies analysed in the current paper. For each asset, the symbol, the name, the category (if available) and the corresponding sector (if available) according to the taxonomy proposed by \cite{Messari} is reported.}
\label{tab:coins_table}\\
\toprule
\textbf{Symbol} & \textbf{Name}                   & \textbf{Category}       & \textbf{Sector}                   \\* \midrule
\endfirsthead
\endhead
\bottomrule
\endfoot
\endlastfoot
AAVE            & Aave                            & Financial               & Lending                           \\
ADA             & Cardano                         & Infrastructure          & Smart Contract Platforms          \\
ADX             & Adex                            & Media and Entertainment & Advertising                       \\
AERGO           & Aergo                           & Infrastructure          & Enterprise and BaaS               \\
ALCX            & Alchemix                        & Financial               & Lending                           \\
ALGO            & Algorand                        & Infrastructure          & Smart Contract Platforms          \\
ALICE           & Alice                           & Media and Entertainment & Gaming                            \\
ALPACA          & Alpaca Finance                  & -                       & -                                 \\
ALPHA           & Alpha Finance                   & Financial               & Asset Management                  \\
AMP             & Amp                             & Financial               & Payment Platforms                 \\
ANT             & Aragon Network                  & Infrastructure          & Misc                              \\
AR              & Arweave                         & Services                & File Storage                      \\
ARPA            & Arpa                            & -                       & -                                 \\
ATOM            & Cosmos                          & Infrastructure          & Smart Contract Platforms          \\
AUDIO           & Audius                          & Media and Entertainment & Content Creation and Distribution \\
AUTO            & Cube                            & Services                & AI                                \\
AVA             & Travala.com                     & Payments                & Payment Platforms                 \\
AVAX            & Avalanche                       & Infrastructure          & Smart Contract Platforms          \\
AXS             & Axie Infinity                   & Media and Entertainment & Gaming                            \\
BADGER          & Badger DAO                      & Financial               & Decentralized Exchanges           \\
BAKE            & BakerySwap                      & -                       & -                                 \\
BAL             & Balancer                        & Financial               & Decentralized Exchanges           \\
BAND            & Band Protocol                   & Infrastructure          & Data Management                   \\
BAR             & FC Barcelona Fan Token          & -                       & -                                 \\
BAT             & Basic Attention Token           & Media and Entertainment & Advertising                       \\
BCH             & Bitcoin Cash                    & Payments                & Currencies                        \\
BEL             & Bella                           & Financial               & Asset Management                  \\
BICO            & Biconomy                        & -                       & -                                 \\
BIFI            & Beefy.Finance                   & -                       & -                                 \\
BNB             & BNB                             & Financial               & Smart Contract Platforms          \\
BNT             & Bancor                          & Financial               & Decentralized Exchanges           \\
BOND            & BarnBridge                      & Financial               & Derivatives                       \\
BTC             & Bitcoin                         & Payments                & Currencies                        \\
BTCST           & Bitcoin Standard Hashrate Token & -                       & -                                 \\
BURGER          & Burger Swap                     & -                       & -                                 \\
CAKE            & PancakeSwap                     & -                       & -                                 \\
CELO            & Celo                            & Infrastructure          & Smart Contract Platforms          \\
CELR            & Celer Network                   & Infrastructure          & Scaling                           \\
CFX             & Conflux Network                 & Infrastructure          & Smart Contract Platforms          \\
CHR             & Chromia                         & Infrastructure          & Application Development           \\
CHZ             & Chiliz                          & Media and Entertainment & Payment Platforms                 \\
CKB             & Nervos Network                  & Infrastructure          & Smart Contract Platforms          \\
COCOS           & Cocos-BCX                       & -                       & -                                 \\
COTI            & Coti                            & Infrastructure          & Application Development           \\
CRV             & Curve                           & Financial               & Decentralized Exchanges           \\
CTK             & CertiK                          & Infrastructure          & Smart Contract Platforms          \\
CTSI            & Cartesi                         & Infrastructure          & Smart Contract Platforms          \\
CTXC            & Cortex                          & Services                & AI                                \\
CVX             & Convex Finance                  & -                       & -                                 \\
DASH            & Dash                            & Payments                & Currencies                        \\
DTA             & DATA                            & Media and Entertainment & Advertising                       \\
DENT            & Dent                            & Services                & Data Management                   \\
DEXE            & DeXe                            & -                       & -                                 \\
DF              & dForce                          & -                       & -                                 \\
DGB             & DigiByte                        & Payments                & Currencies                        \\
DIA             & DIA                             & Infrastructure          & Data Management                   \\
DODO            & DODO                            & Financial               & Decentralized Exchanges           \\
DOGE            & Dogecoin                        & Payments                & Currencies                        \\
DOT             & Polkadot                        & Infrastructure          & Smart Contract Platforms          \\
EGLD            & MultiversX                      & Infrastructure          & Smart Contract Platforms          \\
ELF             & aelf                            & Services                & Shared Compute                    \\
ENJ             & Enjin Coin                      & Media and Entertainment & Gaming                            \\
ENS             & Ethereum Name Service           & Infrastructure          & Identity                          \\
EOS             & EOS                             & Infrastructure          & Smart Contract Platforms          \\
ERN             & Ethernity Chain                 & Media and Entertainment & Collectibles                      \\
ETC             & Ethereum Classic                & Infrastructure          & Smart Contract Platforms          \\
ETH             & Ethereum                        & Infrastructure          & Smart Contract Platforms          \\
FARM            & Harvest Finance                 & Financial               & Asset Management                  \\
FET             & Fetch.ai                        & Infrastructure          & Artificial Intelligence           \\
FIDA            & Bonfida                         & -                       & -                                 \\
FIL             & Filecoin                        & Infrastructure          & File Storage                      \\
FIO             & FIO Protocol                    & Payments                & Interoperability                  \\
FLOW            & Flow                            & Infrastructure          & Smart Contract Platforms          \\
FORTH           & Ampleforth Governance Token     & Payments                & Currencies                        \\
FRONT           & Frontier                        & Infrastructure          & Asset Management                  \\
FTM             & Fantom                          & Infrastructure          & Smart Contract Platforms          \\
FTT             & FTX Token                       & Financial               & Centralized Exchanges             \\
FXS             & Frax Share                      & Financial               & Stablecoins                       \\
GALA            & Gala                            & Media and Entertainment & Gaming                            \\
GHST            & Aavegotchi                      & Media and Entertainment & Gaming                            \\
GRT             & The Graph                       & Infrastructure          & Data Management                   \\
HBAR            & Hedera Hashgraph                & Infrastructure          & Smart Contract Platforms          \\
HIVE            & Hive                            & Media and Entertainment & Content Creation and Distribution \\
HNT             & Helium                          & Infrastructure          & IoT                               \\
HOT             & Holo                            & Infrastructure          & Application Development           \\
ICP             & Internet Computer               & Infrastructure          & Smart Contract Platforms          \\
ICX             & ICON                            & Infrastructure          & Smart Contract Platforms          \\
IDEX            & IDEX                            & Financial               & Decentralized Exchanges           \\
ILV             & Illuvium                        & Media and Entertainment & Gaming                            \\
INJ             & Injective Protocol              & Financial               & Derivatives                       \\
IOST            & IOST                            & Infrastructure          & Smart Contract Platforms          \\
IOTX            & IoTeX                           & Infrastructure          & IoT                               \\
IQ              & Everipedia                      & -                       & -                                 \\
JASMY           & Jasmy                           & -                       & -                                 \\
JOE             & Trader Joe                      & Financial               & Decentralized Exchanges           \\
JST             & JUST                            & Financial               & Decentralized Exchanges           \\
KLAY            & Klaytn                          & Infrastructure          & Smart Contract Platforms          \\
KNC             & KyberNetwork                    & -                       & -                                 \\
KP3R            & Keep3rV1                        & -                       & -                                 \\
KSM             & Kusama                          & Infrastructure          & Smart Contract Platforms          \\
LAZIO           & Lazio Fan Token                 & -                       & -                                 \\
LINA            & Linear                          & -                       & -                                 \\
LINK            & Chainlink                       & Services                & Data Management                   \\
LIT             & Litentry                        & -                       & -                                 \\
LPT             & Livepeer                        & Infrastructure          & Shared Compute                    \\
LRC             & Loopring                        & Financial               & Decentralized Exchanges           \\
LSK             & Lisk                            & Infrastructure          & Application Development           \\
LTC             & Litecoin                        & Payments                & Currencies                        \\
LTO             & LTO Network                     & Infrastructure          & Enterprise and BaaS               \\
MASK            & Mask Network                    & Services                & Data Management                   \\
MATIC           & Polygon                         & Infrastructure          & Scaling                           \\
MBOX            & Mobox                           & -                       & -                                 \\
MC              & Merit Circle                    & Financial               & Gaming                            \\
MDX             & MDX                             & Financial               & Decentralized Exchanges           \\
MINA            & Mina                            & Infrastructure          & Smart Contract Platforms          \\
MIR             & Mirror Protocol                 & Financial               & Derivatives                       \\
MKR             & Maker                           & Financial               & Lending                           \\
MLN             & Enzyme Finance                  & Financial               & Asset Management                  \\
MOVR            & Moonriver                       & Infrastructure          & Smart Contract Platforms          \\
MTL             & Metal                           & Payments                & Payment Platforms                 \\
NEAR            & NEAR Protocol                   & Infrastructure          & Smart Contract Platforms          \\
NEO             & NEO                             & Infrastructure          & Smart Contract Platforms          \\
NMR             & Numeraire                       & Financial               & Asset Management                  \\
NULS            & NULS                            & Infrastructure          & Enterprise and BaaS               \\
OCEAN           & Ocean Protocol                  & Services                & Data Management                   \\
OGN             & Origin Protocol                 & -                       & -                                 \\
OM              & MANTRA DAO                      & Financial               & Lending                           \\
OMG             & OMG Network                     & Infrastructure          & Scaling                           \\
ONE             & Harmony                         & -                       & -                                 \\
ONT             & Ontology                        & Infrastructure          & Smart Contract Platforms          \\
ORN             & Orion Protocol                  & Financial               & Decentralized Exchanges           \\
OXT             & Orchid                          & Services                & Data Management                   \\
PEOPLE          & ConstitutionDAO                 & -                       & -                                 \\
PERP            & Perpetual Protocol              & Financial               & Derivatives                       \\
PHA             & Phala.Network                   & -                       & -                                 \\
PLA             & PLA                             & Payments                & Gaming                            \\
POLS            & Polkastarter                    & Financial               & Crowdfunding                      \\
POND            & Marlin                          & Infrastructure          & Smart Contract Platforms          \\
POWR            & Power Ledger                    & Services                & Energy                            \\
PROM            & Prometeus                       & -                       & -                                 \\
PSG             & Paris Saint-Germain Fan Token   & -                       & -                                 \\
PYR             & Vulcan Forged                   & Media and Entertainment & Gaming                            \\
QNT             & Quant Network                   & Infrastructure          & Interoperability                  \\
QTUM            & Qtum                            & Infrastructure          & Smart Contract Platforms          \\
RAD             & Radicle                         & -                       & -                                 \\
RARE            & SuperRare                       & -                       & -                                 \\
RAY             & Raydium                         & Financial               & Decentralized Exchanges           \\
REEF            & Reef                            & -                       & -                                 \\
REN             & Ren                             & Financial               & Interoperability                  \\
REQ             & Request Network                 & Financial               & Payment Platforms                 \\
RNDR            & Render Token                    & Services                & Shared compute                    \\
ROSE            & Oasis Network                   & -                       & -                                 \\
RSR             & Reserve Rights                  & Financial               & Asset Management                  \\
RUNE            & THORChain                       & -                       & -                                 \\
RVN             & Ravencoin                       & Payments                & Currencies                        \\
SAND            & The Sandbox                     & Media and Entertainment & Gaming                            \\
SC              & Siacoin                         & Services                & File Storage                      \\
SCRT            & Secret Network                  & Infrastructure          & Smart Contract Platforms          \\
SHIB            & Shiba Inu                       & Payments                & None                              \\
SKL             & SKALE Network                   & Infrastructure          & Scaling                           \\
SNX             & Synthetix                       & Financial               & Derivatives                       \\
SOL             & Solana                          & Infrastructure          & Smart Contract Platforms          \\
SPELL           & Spell Token                     & Financial               & Lending                           \\
SRM             & Serum                           & Financial               & Decentralized Exchanges           \\
STMX            & StormX                          & Payments                & Rewards                           \\
STPT            & Standard Tokenization Protocol  & -                       & -                                 \\
STRAX           & Stratis                         & -                       & -                                 \\
STX             & Stacks                          & Infrastructure          & Smart Contract Platforms          \\
SUN             & SUN                             & -                       & -                                 \\
SUPER           & SuperFarm                       & -                       & -                                 \\
SUSHI           & SushiSwap                       & Financial               & Decentralized Exchanges           \\
SXP             & Swipe                           & Financial               & Payment Platforms                 \\
SYS             & Syscoin                         & Infrastructure          & Scaling                           \\
THETA           & Theta Network                   & Media and Entertainment & Content Creation and Distribution \\
TOMO            & TomoChain                       & Infrastructure          & Smart Contract Platforms          \\
TRB             & Tellor                          & Infrastructure          & Data Management                   \\
TRIBE           & TRIBE                           & -                       & -                                 \\
TRX             & TRON                            & Infrastructure          & Smart Contract Platforms          \\
TWT             & Trust Wallet Token              & Financial               & Payment Platforms                 \\
UFT             & UniLend                         & Financial               & Lending                           \\
UNFI            & Unifi Protocol DAO              & Financial               & Interoperability                  \\
UNI             & Uniswap                         & Financial               & Decentralized Exchanges           \\
UTK             & UTRUST                          & Payments                & Payment Platforms                 \\
VET             & VeChain                         & Infrastructure          & Smart Contract Platforms          \\
WAVES           & Waves                           & Infrastructure          & Smart Contract Platforms          \\
WAXP            & WAX                             & Media and Entertainment & Collectibles                      \\
WIN             & WINkLink                        & -                       & -                                 \\
WING            & Wing Finance                    & Financial               & Lending                           \\
WRX             & WazirX                          & Financial               & Centralized Exchanges             \\
XEC             & eCash                           & Payments                & Currencies                        \\
XLM             & Stellar                         & Payments                & Currencies                        \\
XMR             & Monero                          & Payments                & Currencies                        \\
XTZ             & Tezos                           & Infrastructure          & Smart Contract Platforms          \\
XVG             & Verge                           & Payments                & Currencies                        \\
XVS             & VENUS                           & Financial               & Lending                           \\
YFI             & yearn.finance                   & Financial               & Asset Management                  \\
ZEC             & Zcash                           & Payments                & Currencies                        \\
ZIL             & Zilliqa                         & Infrastructure          & Smart Contract Platforms          \\
ZRX             & 0x                              & Financial               & Decentralized Exchanges           \\* \bottomrule
\end{longtable}

\end{document}